\documentclass[twocolumn]{aastex701}

\usepackage{amsmath}
\usepackage{amssymb}
\usepackage{graphicx}
\usepackage{listings}
\usepackage{xcolor}
\usepackage{url}
\usepackage{paralist}
\usepackage{txfonts}
\usepackage{tikz}
\usepackage{color, colortbl}
\usepackage{varwidth}
\usepackage{enumitem}   

\definecolor{pearl}{rgb}{0.94, 0.92, 0.84}
\bibstyle{aasjournal}
\bibpunct{(}{)}{;}{a}{}{,}
\usepackage{txfonts}
\def\tens#1{\ensuremath{\mathsf{#1}}}

\newcommand{\T}           {{^{\rm T}}}
\newcommand{\A}           {{\tens{A}}}
\newcommand{\F}           {{\tens{F}}}
\newcommand{\W}           {{\tens{W}}}

\newcommand{\E}           {{\tens{E}}}

\newcommand{\J}           {{\tens{J}}}

\newcommand{\ostar}       {{\circledast}}

\def\G{{\tens{G}}}
\def\DG{{\tens{D}}}
\def\M{{\bf M}}

\def\JS#1#2#3{{\tens{J}^{#2}_{\!#1}#3}}
\def\MS#1#2#3{{\tens{M}^{#2}_{#1}#3}}
\def\E#1#2#3{{\tens{E}^{#2}_{#1}#3}}

\newcommand\setItemnumber[1]{\setcounter{enumi}{\numexpr#1-1\relax}}
\DeclareRobustCommand{\rchi}{{\mathpalette\irchi\relax}}
\newcommand{\irchi}[2]{\raisebox{\depth}{$#1\chi$}}
\newcommand\myhref[2]{\href{#1}{#2}\footnote{#1}}

\usepackage{hyperref}
\hypersetup{
    colorlinks,
    citecolor=black,
    filecolor=black,
    linkcolor=blue,
    urlcolor=red
}

\definecolor{codegreen}{rgb}{0,0.6,0}
\definecolor{codegray}{rgb}{0.5,0.5,0.5}
\definecolor{codepurple}{rgb}{0.58,0,0.82}
\definecolor{backcolour}{rgb}{0.95,0.95,0.92}
\lstdefinestyle{mystyle}{
    backgroundcolor=\color{backcolour},
    commentstyle=\color{codegreen},
    keywordstyle=\color{magenta},
    numberstyle=\tiny\color{codegray},
    stringstyle=\color{codepurple},
    basicstyle=\ttfamily\footnotesize,
    breakatwhitespace=true,
    breaklines=true,
    captionpos=b,
    keepspaces=true,
    numbers=left,
    numbersep=5pt,
    showspaces=false,
    showstringspaces=false,
    showtabs=false,
    tabsize=2
}
\begin{document}

\title{An Algorithm Architecture for Radio Interferometric
  Data Processing} \shorttitle{Algorithms
  Architecture for Aperture Synthesis Telescopes}
\shortauthors{Bhatnagar et al.}
\correspondingauthor{S. Bhatnagar}
\email[show]{sbhatnag@nrao.edu}

\author{S.~Bhatnagar}
\affiliation{National Radio Astronomy Observatory, 
PO Box O, 1003 Lopezville Rd,Socorro, NM 87801, USA}
\email[show]{sbhatnag@nrao.edu}

\author{U.~Rau}
\affiliation{National Radio Astronomy Observatory, 
PO Box O, 1003 Lopezville Rd,Socorro, NM 87801, USA}
\email[show]{rurvashi@nrao.edu}

\author{M.~Hsieh}
\affiliation{National Radio Astronomy Observatory, 
PO Box O, 1003 Lopezville Rd,Socorro, NM 87801, USA}
\email[show]{mhsieh@nrao.edu}

\author{J.~Kern}
\affiliation{National Radio Astronomy Observatory, 
520 Edgemont Rd, Charlottesville, VA 22903, USA}
\email[show]{jkern@nrao.edu}

\author{R.~Xue}
\affiliation{National Radio Astronomy Observatory, 
520 Edgemont Rd, Charlottesville, VA 22903, USA}
\email[show]{rxue@nrao.edu}

\received{June 24, 2025}
\revised{Aug. 08, 2025}
\accepted{Aug. 19, 2025}

\begin{abstract}
We present a foundational, scalable algorithm architecture for
processing data from aperture synthesis radio telescopes.  The
analysis leading to the architecture is rooted in the theory of
aperture synthesis, signal processing and numerical optimization
keeping it scalable for variations in computing load, algorithmic
complexity, and accommodate the continuing evolution of algorithms.
It also adheres to scientific software design principles and use of
modern performance engineering techniques providing a stable foundation for
long-term scalability, performance, and development cost.

We first show that algorithms for both calibration and imaging
algorithms share a common mathematical foundation and can be expressed
as numerical optimization problems.  We then decompose the resulting
mathematical framework into fundamental conceptual architectural
components, and assemble calibration and imaging algorithms from these
foundational components.

For a physical architectural view, we used a library of algorithms
implemented in the {\tt LibRA} software for the various architectural
components, and used the {\tt Kokkos} framework in the compute-intensive
components for performance portable implementation. This was
deployed on hardware ranging from desktop-class computers to multiple
super-computer class high-performance computing (HPC) and
high-throughput computing (HTC) platforms with a variety of CPU and
GPU architectures, and job schedulers ({\tt HTCondor} and {\tt
  Slurm}).

As a test, we imaged archival data from the NSF's Karl G. Jansky Very
Large Array (VLA) telescope in the A-array configuration for the
Hubble Ultra Deep Field.  Using over 100 GPUs we achieve a processing
rate of $\sim\!\!2$~Terabyte per hour to make one of the deepest
images in the 2 -- 4 GHz band with an RMS noise of
$\sim\!\!1~\mu$Jy/beam.

\end{abstract}

\section{Introduction}
Indirect imaging devices, like an aperture synthesis radio telescope,
record the data in the Fourier domain, and the raw data needs to be
Fourier transformed to convert it into an image.  These devices sample
the Fourier domain incompletely and irregularly, and the data is
corrupted by the troposphere and ionospheric layers of the Earth's
atmosphere, and by the chain of imperfect electronic components used
in such instruments. Noise-limited imaging with such telescopes
requires computationally- and I/O-intensive algorithms to mitigate
artifacts arising from incomplete and irregular sampling, the presence
of radio frequency interference, and from distortions introduced by
atmospheric effects and instrumental imperfections.

To achieve the scientific requirements of high sensitivity and angular
resolution, modern aperture synthesis radio telescopes utilize
wide-band receivers on hundreds to thousands of antenna elements
spread across distances exceeding 1000 kilometers
\citep[e.g. see][]{ngvla-refdesign}.  The sky brightness distribution,
especially at low radio frequencies, is also stronger and more
wide-spread, requiring imaging over wider field of view which further
increases the data volume and computing load. The resulting data rates
and estimated size of computing for typical imaging experiments with
such telescopes are on the order of 100 terabyte per hour, and tens of
peta floating-point operations per second (PFLOP/s)
\citep{ngVLA-SofC}, respectively.  Achieving this level of performance
necessitates the use of parallel computing across thousands of
Execution Points (EPs), where each EP is defined as a ``compute node''
comprising of multi-core CPUs and GPUs.  These systems typically rely
on heterogeneous computing hardware, incorporating a range of CPU and
GPU architectures.  Consequently, software designed for such platforms
must be both computationally efficient at the EP level and scalable
across large, {\it heterogeneous} computing environments.  Imaging
with such telescopes therefore fundamentally requires large scale
heterogeneous parallel computing.  While this has been recognized in
the radio astronomy (RA) community for sometime, routine use of the
available computing resources has largely remained elusive, even
though multiple computing platforms with the requisite computing power
are available.

Assessments from the computing industry and the high-performance
computing (HPC) community indicate that computing hardware
architectures are expected to evolve more rapidly over the coming
decade than in previous ones \citep[e.g. see][]{PlentyOfRoomAtTheTop}.
The resulting gains in runtime performance will primarily benefit
computing workflows capable of efficiently parallelizing across
multiple scales to exploit heterogeneous collections of massively
parallel and multi-core hardware.  Consequently, harnessing the full
computational potential of future systems requires the development of
algorithms that are highly parallelizable and implementations that are
performance-portable scaling effectively from intra-node parallelism
(e.g., across multi-core CPUs and multiple GPUs within a single EP) to
inter-node parallelism (e.g., across thousands of EPs in tightly
coupled or distributed computing environments).

In the RA domain, the advent of next-generation high-sensitivity,
high-resolution telescopes is expected to drive a corresponding
acceleration in the evolution of algorithms.  These algorithms must
deliver improved imaging performance while being optimized for
execution on rapidly evolving, heterogeneous computing architectures,
and must be designed to scale efficiently across multiple levels of
parallelism.  The overall computing landscape comprising of new EP and
parallel computing architectures, and new domain algorithms for the
future therefore remains unpredictable.

Future-readiness in this rapidly evolving landscape over the life time
of the telescopes, and the software stability necessary on even longer
time scales for creative research in algorithm development requires an
algorithm architecture grounded in foundational principles.  This is
essential not only for reusability and scalability, but even more
critically for enabling the integration of future, generally
applicable algorithms (as opposed to bespoke algorithms narrowly
tailored to specific use cases).  A software system based on such an
architecture is expected to scale in an evolving landscape of
computing technology, and domain requirements of increasing data
volume, complexity, or computational demands, without requiring
frequent time-consuming and expensive redesign and re-implementation
of the domain code.

In this paper, we first develop a theoretical framework grounded in
the fundamental principles of aperture synthesis and signal
processing, and demonstrate that existing algorithms for both
calibration and imaging can be described as numerical optimization
problems and fit in the {\it same} framework.  We then present a
component-level decomposition by identifying a set of foundational
algorithmic components, and construct a hierarchical Algorithms
Architecture based on these elements.  In this architecture, core
algorithms are composed from foundational components, while
higher-level, domain-specific schemes and processing pipelines are
systematically constructed from these core algorithms.  As an
illustrative example, we show that a core set of domain algorithms
serving as the fundamental building blocks for advanced data
processing workflows can be effectively developed within this
architectural framework.

We validate several aspects of the architecture through a concrete
example.  An accompanying paper presents the physical
architectural view and demonstrates the use of modern performance
engineering tools to achieve a performance-portable implementation.
This implementation is deployable across a wide range of hardware
platforms, from laptops and desktop systems to distributed wide-area
networks, such as the Open Science Grid (OSG) in the U.S. and
super-computing clusters with tightly coupled nodes. We also discuss
the design of a software library comprising of reusable modular
components, along with their optimization and runtime performance on
heterogeneous platforms with a diverse set of CPU and GPU
architectures.

\section{Theoretical foundation}
\label{Sec:Theory}
The measurements made with a radio interferometric array telescope is
described with a linear measurement equation (ME) parameterized for
the contributions of the sky brightness distribution, the telescope
optics and the electronic elements to the total signal.  In the
description below, we borrow the theoretical formulation from a number
of publications (\cite{HBS1,HBS2}, \cite{Rau_ieee}, \cite{NumRecipes},
\cite{Bhatnagar_2017_PSC}).  The analysis here is limited to
calibrating the data and imaging the sky radio brightness
distribution.  The ME can be expressed as
\begin{eqnarray}
\label{Eq:ME}
V^{Obs}_{ij}(\nu) &=&
\MS{ij}{DI}{(\nu)}\F\left[\MS{ij}{DD}{(\vec{s},\nu)}~I(\vec{s},\nu)\right]
+ n_{ij}(\nu)
\end{eqnarray}
$V^{Obs}_{ij}$ is the full-polarization data vector, referred to as
the {\it visibilities}, from a single interferometer made with a pair
of antennas indexed by subscripts $i$ and $j$.  $I$ represents the
full-polarization sky brightness distribution vector in the direction
$\vec{s}$ and at frequency $\nu$.  $\MS{ij}{DI}{}$ and $\MS{ij}{DD}{}$
are the direction independent (super-script $DI$) and the direction
dependent (super-script $DD$) Mueller matrices respectively.  These
describe the mixing of the elements of the full-polarization vector
for the interferometer $ij$. $\F$ is the Fourier transform operator,
$n_{ij}$ represents the measurement noise which has contributions all
statistically significant emission not represented by $I$, Earth's
atmosphere and the telescope hardware.  The explicit frequency
dependence is dropped in the treatment below for notational clarity,
except where necessary to show that the architecture covers algorithms
that account for frequency dependence (e.g. by treating and
aggregating data binned in narrow frequency ranging in a variety of
ways).  Detailed description of these can be found elsewhere (e.g. see
\cite{Rau_ieee} and references therein).

The Mueller matrices are $4\times 4$ matrices which in most cases can
be factored into antenna-based $2\times 2$ Jones matrices as
\begin{eqnarray}
\MS{ij}{DI}{} &=& \J^{DI}_{\!i}\otimes \J^{DI^\dag}_{\!j}\\
\MS{ij}{DD}{} &=& \J^{DD}_{\!i}\ostar \J^{DD^\dag}_{\!j}
\end{eqnarray}
The symbols $\otimes$ and $\ostar$ represent outer product and outer
convolution\footnote{The element-by-element algebra of the
outer-convolution operator is the same as that of the outer-product
operator used in the direction-independent (DI) description by
\cite{HBS1}, except that the complex multiplications of outer-product
are replaced by the convolution operator \citep{WBAWProjection}.}
operations respectively.  Solvers directly solve for $\J_i$s, and
$\MS{ij}{}{}$ are constructed for use in calibration and imaging.

The {\it unknowns} in Eq.~\ref{Eq:ME} are the elements of the Jones
matrices ($\J_i$, or of $\MS{ij}{}{}$ in general) and the sky
brightness distribution ($I$).  The primary goal of the calibration
process is to solve for $\MS{}{DI}{}$ and $\MS{}{DD}{}$ to a
sufficient accuracy to allow solving for $I$.  Solving for $I$ to a
level statistically consistent with $n_{ij}$ is the primary goal of
the imaging process. In many real-life use cases these unknowns are
further parameterized and the solvers directly solve for these
implicit parameters.

The only {\it known} quantities in Eq.~\ref{Eq:ME} are the observed
data ($V^{Obs}$) and the noise model.  The noise term can be described
only statistically and it can be shown (and measured) to be normally
distributed.  Any successful algorithm for calibration and imaging
therefore must lead to solutions for $\MS{ij}{}{}$s and $I$ that
satisfies Eq.~\ref{Eq:ME} and leaves residuals that are statistically
consistent with $n_{ij}$.  That is, the probability distribution of
residuals $V^R = V^{obs} - V^{Mod}$ must be consistent with the
distribution of $n_{ij}$ where $V^{Mod}$ is written in terms of the
solutions $\MS{ij}{Mod}{}$ and $I^{Mod}$ as
\begin{eqnarray}
\label{Eq:ME-MOD}
V^{Mod}_{ij} &=&
\MS{ij}{DI,Mod}{}\F\left[\MS{ij}{DD,Mod}{}(\vec{s})~I^{Mod}(\vec{s})\right]
\end{eqnarray}

The calibration and imaging algorithms can therefore be described as
algorithms to estimate $n_{ij}$ for the given $V^{Obs}$ by
appropriately selecting $\MS{ij}{}{}$ and $I$ from an otherwise
unbounded parameter space. A successful algorithm therefore requires
a metric which when minimized (within the bounds of algorithmically
determined constraints on the search space), leaves residual
consistent with the noise.  It can be shown that $n_{ij}$ is normally
distributed, for which the weighted $\rchi^2$ function is the optimal
metric -- a standard result of statistical analysis -- given by
\begin{eqnarray}
\label{Eq:Chisq}
\rchi^2 &=& \sum_{ij}\left[V^R\right]^\dag_{ij}w_{ij}\left[V^R\right]_{ij}
\end{eqnarray}
where $w_{ij}$ are the measurement weights proportional to the inverse
of the variance in the measurement noise ($n_{ij}$).

For runtime scaling, the FFT algorithm is used for the $\F$ operator
in Eq.~\ref{Eq:ME-MOD}, which requires data on a regular
grid. $V^{Mod}_{ij}$ however exists only for coordinates determined by
the location of the antennas $i$ and $j$ on the ground, which are not
on a regular grid.  Therefore, in practice a re-sampling operator
$\DG$ is required to resample the output of the FFT to the coordinates
of the sampled data.  Here, this operator is parameterized as
$\DG(c_{ij})$, where $c_{ij}$ is the interpolation kernel used to
compute $V^{Mod}_{ij}$ as
\begin{eqnarray}
\label{Eq:ME-MOD-GRIDDED}
V^{Mod}_{ij} &=&
\MS{ij}{DI,Mod}{}\DG(c_{ij})\F\left[\MS{ij}{DD,Mod}{}(\vec{s})~I^{Mod}(\vec{s})\right]
\end{eqnarray}
Similarly, the reverse operator $\G(c_{ij})$ is required for
resampling the irregularly sampled data (e.g. $V^R$) on to a regular
grid and compute the image using the FFT algorithm for Fourier
transform.
\begin{eqnarray}
\label{Eq:IMAGE-MOD}
I^{Mod} &=& \F^{\dag}\sum_{ij}\G(c_{ij})w_{ij}V^{R}(\vec{u}_{ij})
\end{eqnarray}
where $\vec{u}_{ij}$ is the location of the interferometer $ij$ on the
spatial frequency plane, and the summation indicates accumulation on a
grid at the pixel corresponding to $\vec{u}_{ij}$.  For brevity,
$V_{ij}$ is used to indicate measurements at $\vec{u}_{ij}$ in
the spatial frequency domain.  Symbols $\DG_{ij}$ and $\G_{ij}$ are
used where $c_{ij}$ varies with $i$ and $j$, and just $\DG$ and $\G$
otherwise.  These in the RA literature are also referred to as the
``de-gridding'' and the ``gridding'' operators, respectively. The
convolutional kernels used in these operators is also explicitly
specified as a parameter only where it is necessary for clarity (as in
Sec.~\ref{Sec:PhysicalArch}).

The parameter space in which the $\rchi^2$ surface is defined can be
thought of as a space of calibration and image models (the
$\MS{ij}{Mod}{}$ and $I^{Mod}$).  Minimizing $\rchi^2$ using the
derivative to estimate the update direction in this space for
iteratively updating the models leads to directed-search algorithms.
We show below that most existing algorithms can be described in this
framework.  For example, we show that for iterative image
reconstruction algorithms, the residual image is the update direction,
which is used to iteratively update the image model.  Similarly,
equating to zero the derivative with respect to $\J_i$s (or its
elements) gives the expressions for the update direction in core
calibration algorithms. These algorithms have proved to be both
computationally and numerically most competitive and are the most
commonly used algorithms.  Other variants that differ in how the
constraints are imposed to limit the search space are also possible.
Non-directed search algorithms (e.g., the family of Monte Carlo-style,
or machine learning based algorithms) are also admissible and some are
in use.  Note that even for such algorithms, goodness-of-fit
criteria unsurprisingly involves the use of the $\rchi^2$ function, and
in many cases even its derivative.

\begin{figure}
\begin{center}
  \includegraphics [height=2.0in]{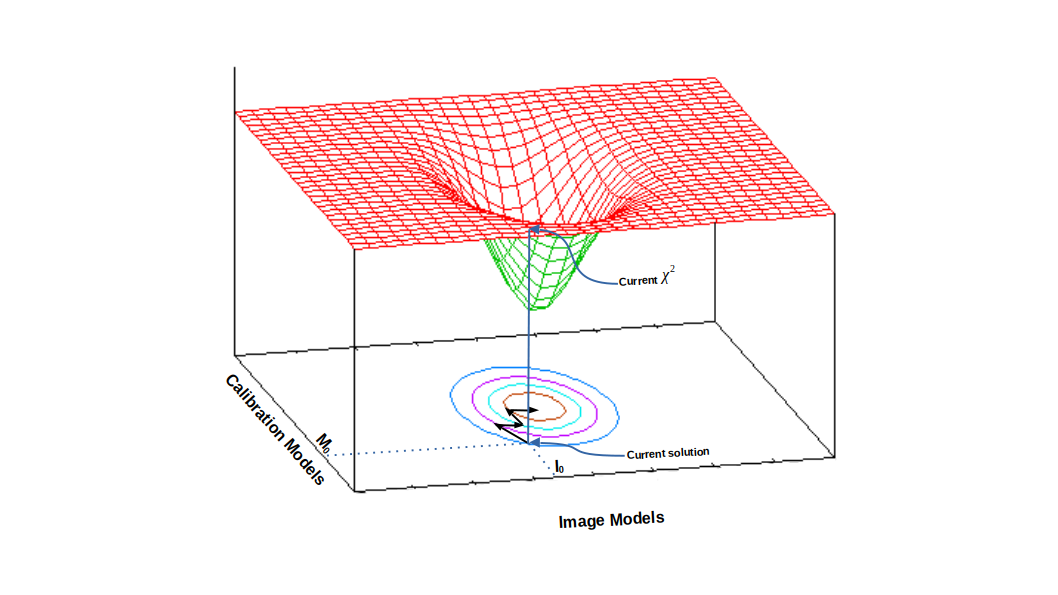}
  \caption {\small The $\rchi^2$ hyper-surface (in red) shown in a
    space of calibration and image models. The current solution is
    shown as the point with coordinates $(\MS{o}{}{},I_o)$, where $\MS{o}{}{}$ and
    $I_o$ represent the starting calibration and image model
    respectively.  The arrows overlayed on the contour plot show
    successive application of the calibration and imaging solvers to
    iteratively move the solution to minimize the metric in
    Eq.~\ref{Eq:Chisq}.}
  \label{Fig:HyperPlane}
\end{center}
\end{figure}

The algorithms for calibration and imaging are iterative in
nature. This is also rooted in the numerical analysis theory
independent of any domain specific knowledge of the parameters of
interest -- while the calibration problem is inherently non-linear in
antenna-based parameters, imaging process constitutes an ill-posed
problem.  Successful algorithms use simplifying assumptions that allow
treating the phase space as locally orthogonal and iteratively moving
the current solution towards the minima in a fundamentally
non-orthogonal search space.  As a result, while the solvers for
  $I^{Mod}$ and $\MS{}{Mod}{}$ are themselves iterative, the solver
  for the full ME (Eq.~\ref{Eq:ME}) in the hyper-space of calibration
  and image models is also iterative in nature.  This is depicted
graphically in Fig.~\ref{Fig:HyperPlane} for conceptual understanding.

No domain-specific knowledge about the parameters of interest has been
used so far.  The ME is derived from the basic theory of aperture
synthesis, and the conclusions about the solvers are based on the
fundamental statistics and signal processing theories.  Solvers for a
parameterized ME based on this analysis are therefore foundational in
nature and would form an important part, if not the basic building
blocks of useful algorithms and data processing workflows.
Architectures based on such a foundational framework will also be
extensible -- they can accommodate new, general-purpose algorithms
that may not yet exist, ensuring long-term adaptability to innovations
in domain algorithms, computing software and hardware technology.

\begin{figure}
\begin{center}
  \includegraphics [height=1.8in,width=3.4in]{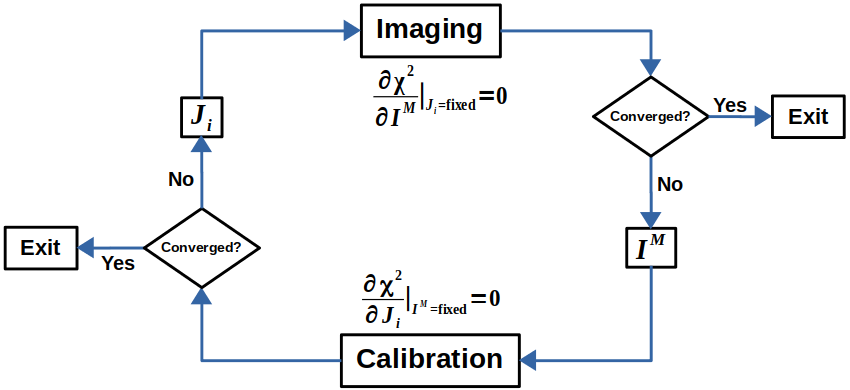}
  \caption {\small The general structure of an iterative
      calibration-imaging solver (also known as the {\it Self
        Calibration} procedure). As an example of hierarchical
      algorithm development, Fig.~\ref{Fig:SelfCal} shows such a
      solver built using the independent calibration and image solvers
      as architectural components developed in Sec.~\ref{Sec:AV-CALIB}
      and \ref{Sec:Imaging} respectively.}
  \label{Fig:CalIm}
\end{center}
\end{figure}

In the following sections we give a brief overview of some relevant 
numerical minimization algorithms and their
decomposition into generic fundamental components.  We then cast the
RA domain core algorithms as function optimization algorithms,
which are then assembled with these generic components.  Higher level
domain algorithms (e.g. the calibration-imaging ``SelfCal loop'' shown
in Fig.~\ref{Fig:CalIm}) and pipelines are assembled from these
foundational domain algorithms in a hierarchical algorithms
architecture.

\section{General structure of the calibration and imaging algorithms}
\label{Sec:GEN-STRUCTURE}
As mentioned earlier, the unknowns in Eq.~\ref{Eq:ME} are the
calibration models $\MS{}{DI}{}$ and $\MS{}{DD}{}$ and the sky
brightness model $I(\vec{s})$.  The $\MS{}{}{}$s model the corruptions
in the data domain, while $I$ models the sky brightness distribution.
Solvers search for optimal models in the hyper-space of calibration
and image models by minimizing the $\rchi^2$ metric subject to
constraints that limit the search space, depicted graphically in
Fig.~\ref{Fig:HyperPlane}.

The coordinates $(M_o, I_o)$ in Fig.~\ref{Fig:HyperPlane} indicates
the location of the current solution.  Ideally, a joint solver that
{\it simultaneously} solves for both models is required to reach the
minimum on this surface.  While such a solver can be assembled using
the basic architectural components derived from numerical optimization
considerations as described in Sec.~\ref{Sec:Components}, the
computational and algorithmic complexity of such solvers is
significantly higher compared to calibration and image solvers
separately.  A widely used approach in numerical optimization
therefore is to assume that the search space is {\it at least} locally
orthogonal and devise schemes to approach the minima by the
application of calibration or imaging algorithms separately in success
steps to move the solution along one of the axis at a time, as shown
with arrows overlayed on the contour plot in
Fig.~\ref{Fig:HyperPlane}.

Calibration algorithms are derived by solving the following system of
equations, keeping $I^{Mod}$ constant
\begin{equation}
\label{Eq:dM}
\left.\frac{\partial \rchi^2}{\partial \JS{i}{Mod}{}}\right|_{I^{Mod}=constant}=0
\end{equation}
Similarly, imaging algorithms are derived keeping the calibration model
$\MS{ij}{Mod}{}$ fixed and solving the system of equations given by
\begin{equation}
\label{Eq:dI}
\left.\frac{\partial \rchi^2}{\partial I^{Mod}}\right|_{\MS{ij}{Mod}{}=constant}=0
\end{equation}
Solvers for these system of equations are iterative themselves, and
can be assembled with specific implementations of components used in
numerical optimization as described in sections~\ref{Sec:Components}
and \ref{Sec:ARCH-VIEWS}.

A joint solver can be assembled with successive application of these
calibration and imaging algorithms as shown in Fig~\ref{Fig:CalIm}.
The iterations are bootstrapped, e.g. with a pre-existing image model
to derive improved calibration models, which are then used to further
improve the image model at each iteration.  This general structure is
referred to as the Self-Calibration procedure and embodies the
assumption of a locally orthogonal phase space.

Higher level, more complex algorithms can thus be built by assembling
lower level simpler components in a hierarchical manner.  Calibration
and imaging algorithms are assembled using foundational components,
while a joint solver for the full ME is assembled using calibration
and imaging algorithms.  Even higher level constructs, e.g. algorithms
that partition the image or the data, or both, heuristics for
automatic processing pipelines, etc. can be assembled with a
mix-and-match of algorithms and even basic components from lower
levels of the algorithmic hierarchy.

\subsection{Cross domain collaborations}
The fundamental conceptual components required for both calibration
and imaging are the same -- namely, component to (a) compute the
derivative of the $\rchi^2$, (b) compute the residual vector for the
current model parameters, and (c) update the current parameter values.
These steps typically involve some domain knowledge since the models
are domain specific by definition, but also because these steps have
high computational complexity.  For example, calculations of the
residuals for calibration involves solving and application of current
calibration models to compute nominally corrected data (see
Sec.~\ref{Sec:Cal}), while for imaging the derivative computation
involves the expensive data resampling and FFT operations (see
Sec.~\ref{Sec:Imaging}).  Such challenges of overall complexity are
common across other domains that rely on numerical optimization, and
addressing them will benefit from collaborations with experts from
other communities -- particularly in high-performance computing and
numerical optimization communities.  Such cross-domain collaborations
to be effective necessitates that the problem formulation is expressed
in a shared, standard mathematical language -- as against in
non-standard domain-specific narrative.  Casting RA algorithms in the
standard language of numerical optimization therefore offers at least
the following benefits:
\begin{enumerate}
\item Enables participation by non-RA experts.  People from many other
  fields who are fluent in numerical techniques but may not be
  familiar with RA jargon can understand the algorithms and be able to
  contribute effectively.
\item Enables use of well-regarded reliable numerical libraries.
  This, apart from getting RA community plugged into the larger
  community of scientific computing professionals, it also opens up
  the path for experimenting with more sophisticated numerical
  techniques, implementations of which exists with plugable
  components.
\end{enumerate}

{\it The take-away point here is that the structure of both
  calibration and imaging algorithms is mathematically the same, and
  can be described in standard terminology.  Both type of algorithms
  are either direct $\rchi^2$-minimization algorithms, or $\rchi^2$
  minimization forms an integral part of the solver.  An hierarchical
  architecture where domain-specific functionality is built from more
  basic components, which are themselves described in a standard
  mathematical language enables effective cross-domain collaborations
  without the need for the experts from other domains to first fully
  understand domain specific details.
}

\section{Structure of $\rchi^2$ minimization algorithms}
\label{Sec:MINIMIZATION-ALGOS}
This section contains an overview of the numerical optimization
concepts used in later sections (see standard text books,
e.g. \cite{NumRecipes}, for a more comprehensive description).

Starting from an initial estimate, the parameter vector $\vec{P}$ is
updated in the $k^{th}$ iteration as:
\begin{eqnarray}
  \vec{P}^{k+1} &=& \vec{P}^k - {\bf H}^{-1}
                f\left(\vec{\nabla}_{P} \rchi^2\right)
\label{Eq:GENERIC-UPDATE}
\end{eqnarray}
where ${\bf H}$ is the Hessian matrix. Its elements are 
given by
\begin{equation}
H_{ij} = \frac{\partial^2 \rchi^2}{\partial p_{\!i}\partial p_{\!j}}
\end{equation}
where  $p_i$ are elements of $\vec{P}$.

Conceptually, $f\left(\vec{\nabla}_P\rchi^2\right)$ represents the
estimated direction towards the minima on the $\rchi^2$-surface along
which the parameters are moved, and the elements of ${\bf H^{-1}}$
represents the step size.  A pure diagonal ${\bf H}$ would indicate a
strictly orthogonal parameter space, where the update equation
simplifies to:
\begin{eqnarray}
  P_i^{k+1} &=& P^k_i - \left[\frac{\partial^2 \rchi^2}{\partial
      P_{\!i}^{k^2}}\right]^{-1} f\left(\frac{\partial
    \rchi^2}{\partial P_{\!i}^k}\right)
\end{eqnarray}
The residuals, especially closer to the minima, are dominated by noise
and the derivative also approaches small values. The
$\left[H_{ii}\right]^{-1}$ therefore progressively becomes numerically
unstable as the iterations proceeds towards convergence.  Numerical
computation of $H_{ii}$ can also be expensive.  To avoid these
problems, a commonly used solution in general is to use a fixed step
size (e.g. the loop-gain), or an empirically determined value, or a
value derived with domain-specific heuristics.  Using $\alpha$ as the
step size, the update equation is expressed in a familiar form as
\begin{eqnarray}
  \label{Eq:ModelUpdate}
      P_i^{k+1} &\approx& P_i^k - \alpha f\left(\frac{\partial
        \rchi^2}{\partial P_{\!i}^k}\right)
\end{eqnarray}
Speed of convergence is sometimes improved using $H_{ii}$ when
$\rchi^2$ is large (indicating the current solution is far from the
minima) and a constant closer to the minima.  Variants of
Eq.~\ref{Eq:ModelUpdate} used in some other domains are parameterized
with heuristics to update the parameters.  Eqs.~\ref{Eq:SD+GN} and
\ref{Eq:LM} below show two widely used variants.
\begin{quote}
  \item Steepest descent + Gauss-Newton:
\begin{equation}
  \label{Eq:SD+GN}
  P_i^{k+1}  \approx P_i^k - \alpha~f\left(\frac{\partial
    \rchi^2}{\partial P_{\!i}^k}\right) - \beta~{\bf
    h}^{-1}
\end{equation}

\item Levenburg-Marquardt (LM):
\begin{equation}
  \label{Eq:LM}
  P_i^{k+1}  \approx P_i^k - \beta~{\bf
    h^\prime}^{-1}
\end{equation}
\end{quote}
where $h_{ij}^\prime = h_{ij}\left(1+\delta_{ij}\lambda\right)$,
$h_{ij} = \frac{\partial \rchi^2}{\partial p_{\!i}}\frac{\partial
  \rchi^2}{\partial p_{\!j}}$, and $\delta_{ij}=1$ for $i=j$ and zero
otherwise.  Use of Eq.~\ref{Eq:SD+GN} with large $\alpha$ relative to
$\beta$ when the solutions are far from optimal, and increasing the
value of $\beta$ relative to $\alpha$ nearer the minima may
improve convergence and even solutions at low SNR.  Some schemes switch to
Eq.~\ref{Eq:LM} closer to the minima.

Exit criteria to terminate the iterations is often based on
domain-specific heuristic derived from the current value of $\vec{P}$,
the $\rchi^2$ and $\vec{\nabla}_P\rchi^2$, or simply by limiting the
maximum number of iterations.  For example, one of the criteria used
to terminate iterations in image reconstruction is when
$max\left(\vec{\nabla}_P\rchi^2\right) < threshold$ is satisfied.
Calibration iterations are typically terminated when the change in
$\vec{P}$ from one iteration to another is smaller, or when the
$\rchi^2$ is smaller than a threshold.  In the description below we
use $T\left(\vec{P}, \vec{\nabla}_P\rchi^2, \rchi^2\right)$ to
encapsulate the termination heuristics.

{\it An architecture for RA algorithms based on this generic formulation
  would be algorithmically extensible.  Current implementations, which
  are mostly Steepest Descent methods ($\beta=0$ in
  Eq.~\ref{Eq:SD+GN}), can be easily realized while leaving the option
  of using other optimization methods \citep[e.g.,][]{GC-Clean}.  An
  architecture build on fundamentals -- as against on non-standard
  domain-specific narrative -- can also be a foundational basis for
  developing new algorithms which may evolve with computing
  technology.}

\begin{figure*}[t]
\begin{center}
  \includegraphics [height=1.9in,width=\textwidth]{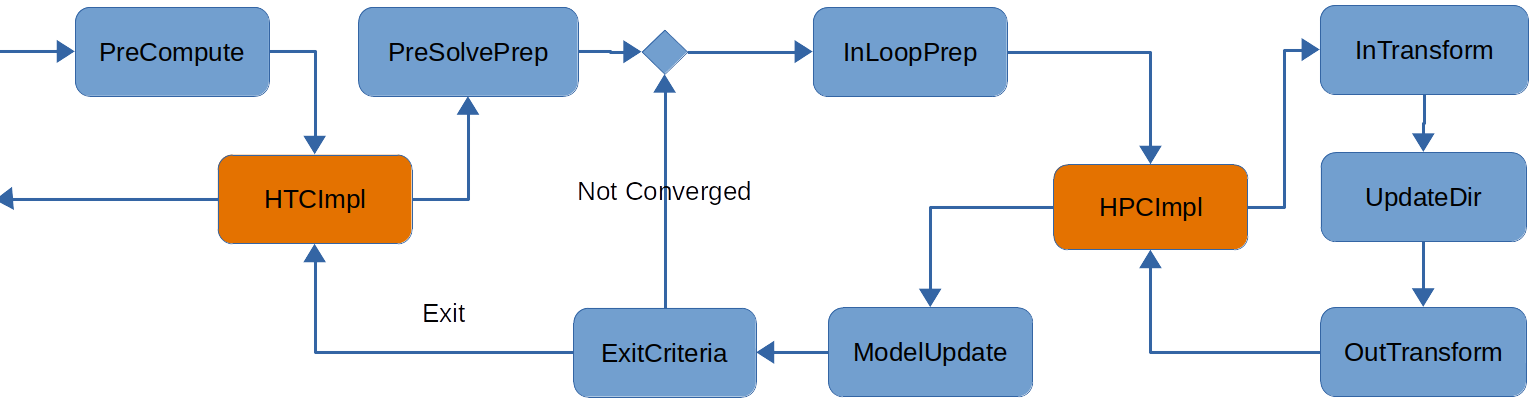}
  \caption {\small The conceptual architecture for a general solver
    built with architectural components derived from
    theoretical analysis.  The components for parallelization or data
    preparation are for optimization and extensibility.}
  \label{Fig:GenSolver}
\end{center}
\end{figure*}

\subsection{Component level decomposition}
\label{Sec:Components}
We decompose the above generic framework into basic, fundamental
components required to develop a foundational architecture for a
generic solver (Fig.~\ref{Fig:GenSolver}).  In
Sec.~\ref{Sec:ARCH-VIEWS}, RA algorithms are also described as
numerical optimization problems, and mapped to the generic solver as
specialized architectural views.

Some of the components are foundational in nature derived from
fundamental theory (e.g. the {\tt UpdateDir} and {\tt ModelUpdate}
components) while some others are derived from the requirements of
extensibility, robustness, runtime reliability and computational
efficiency. These are also expressed in general conceptual terms at
the architectural level (e.g. the {\tt PreSolvePrep} or the {\tt
  InTransform} components).
\begin{enumerate}
\item {\tt UpdateDir:} A component that encapsulates the computation
  of the update direction -- the function $f$ in
  Eq.~\ref{Eq:GENERIC-UPDATE}.  In general, this component is applied
  to all of the data necessary for the required accuracy in the
  solutions and therefore often drives the overall computational
  cost.

\item {\tt ModelUpdate:} A component that encapsulates parameter
  update. Among other domain specific things, this component needs
  access to the derivative, value of the penalty function ($\rchi^2$
  with any additional terms, e.g. various forms of algebraic
  regularization terms), and the current parameter vector.

\item {\tt StepSize:} A sub-component used in the {\tt ModelUpdate}
  component for the step-size to update the vector of parameters in
  the direction given by the {\tt UpdateDir} component.  Introduction
  of this architectural component opens the path for more
  sophisticated approaches that may deliver improved rate of
  convergence, improved numerical performance, or both.

\item {\tt ExitCriteria:} A component that determines the termination
  condition for iterative algorithms.  Implementations typically rely
  on a combination of theoretical guidelines and domain-specific
  heuristics.  These heuristics may require access to the current
  value of the penalty function, its derivatives, parameter vectors,
  the number of iterations executed, and other user-defined parameters
  to ensure robust and reliable termination.
\end{enumerate}

The physics of the instrument and the hardware implementation of it
impose some characteristic scales of coherence on the variations of
the model parameters (e.g. instrumental response as a function of
frequency and polarization, antenna and dish design, and the signal
path through the chain of electronic components, etc.).  The
parameters may vary at a statistically significant level only at
scales larger than these fundamental scales, and are stable at shorter
scales.  It is therefore often both computationally and numerically
advantageous to average the data on these scales. The data may also be
transformed, or normalized for improved numerical performance.  There
may be other operations on the data, e.g., data pre-conditioning,
flagging, selection of only the pertinent data, etc.  Abstracting
these kind of operations away from the fundamental theoretical blocks
increases the re-usability and scalability.  We therefore introduce
the following components in the architecture:
\begin{enumerate}
  \setItemnumber{5}
\item{\tt PreCompute:} A component for caching computations that
  remain fixed for the life cycle of the solver but require
  access to the full data.  For example, shaping the telescope Point
  Spread Function by data reweighing, or caching the kernels used in
  the $\G$ and $\DG$ sub-components of the {\tt UpdateDir} component
  (see Figure~\ref{Fig:ImSolver}, Section~\ref{Sec:Imaging}).

\item {\tt PreSolvePrep:} A component responsible for data preparation
  and curation tasks that are independent of the iterations and may be
  done in parallel on portions of the full data.  Typical operations
  include averaging data along various axes, especially where such
  pre-processing is required prior to the application of the solvers
  for antenna-based parameters (e.g., see Eq.\ref{Eq:VAvg} of
  Sec.\ref{Sec:AV-CALIB}). The placement of this component within the
  overall framework may vary depending on the specific use case.
\item {\tt InLoopPrep:} A component for iteration-dependent data
  transformation and curation operations (e.g. use of data flagging
  information discovered during iterations).
\end{enumerate}

In certain imaging use cases, the same input data contributes to
different outputs (for example, in multi-term multi-frequency
synthesis (MT-MFS) \citep{MSMFS}).  Similarly, there are scenarios
where multiple outputs are generated from input data that has
undergone complex transformations, and these intermediate outputs are
subsequently combined in non-trivial ways to construct the final
desired output. Examples of such use cases include multi-field
imaging, image-plane mosaicking \citep{NRAO_LECTURES}, and wide-field
imaging algorithms such as faceting and the W-projection/W-snapshot
methods \citep{TIM_N_RICK_1992_FACETING, WSnapshot}.  To cover such
use-cases we introduce the following transform components at the input
and/or the output of the {\tt UpdateDir} component:
\begin{enumerate}
  \setItemnumber{8}
\item {\tt InTransform:} A component that applies transform on the 
  input data stream before it is used for the derivative calculation.

\item {\tt OutTransform:} A component that applies transform, including
  aggregation, at the output of the {\tt UpdateDir} component.
\end{enumerate}
These components may have a coupling.  For example, {\tt OutTransform}
component may need to know about the {\tt InTransform} component for
mathematical consistency.  These components may be at the input and
output of the {\tt HPCImpl} component (see
Sec.~\ref{Sec:PARALLELIZATION}), or be one of the sub-components that
together realize the calculations of the derivative in the {\tt
  UpdateDir} component.

Figure~\ref{Fig:GenSolver} shows a block-level diagram for a generic
solver built from these component blocks.  In
Sec.~\ref{Sec:ARCH-VIEWS} we map components of RA algorithms to these
generic components and show that the architecture delivers the
existing algorithms, and is extensible for new algorithms in the
future.

\subsubsection{A comment on parallelization}
\label{Sec:PARALLELIZATION}
There are two components in the block diagram in
Fig.~\ref{Fig:GenSolver} for parallelization -- the {\tt HTCImpl} and
the {\tt HPCImpl} components -- each of which are envisioned to serve
a different and a distinct purpose.  {\tt HTCImpl} should be thought
of as a mechanism for cluster-level parallelization (Same Program
Multiple Data (SPMD) paradigm). {\tt HPCImpl} is a lower-level
parallelization mechanism and may be considered as targeting
node-level, and more specifically, {\it local}, optimization at the EP
level.  For example, it enables the utilization of multiple compute
units (GPUs, CPUs, FPGAs, etc.), multiple cores across CPUs,
partitioned GPUs, and non-trivial coordination between the host CPU
and associated GPU resources on a single EP.  This provides a
framework block for a different class of parallel code that may
require hardware-specific implementations to fully exploit local
hardware and data proximity.  This component may be part of the {\tt
  UpdateStep} block or placed outside it, depending on the overall
runtime optimization strategy.  Its location may also be a
higher-level packaging or API decision, which can be deferred to a
later stage of development.

Put another way, {\tt HTCImpl} represents implementation of High
Throughput Computing (HTC) while {\tt HPCImpl} represents
implementation of High Performance Computing (HPC).  Since multiple
definitions exists, to clarify what we mean by these terms, below is
the definition we use:
\begin{quote}
{ High-throughput computing (HTC) refers to parallel computations
  where individual tasks do not need to interact at runtime, or
  require small bits of information exchange. High-performance
  computing (HPC) on the other hand refers to parallel computations
  where frequent and rapid exchanges of intermediate results may be
  required to perform the calculations. HPC codes are based on tightly
  coupled MPI, OpenMP, GPU, and hybrid programs and require low
  latency interconnect.  In contrast to HPC, which uses a
  single "supercomputer," HTC distributes tasks over many computers
  and collecting results at the end of all parallel tasks.  For example,
  computing grids composed of heterogeneous resources (clusters,
  workstations, etc).  }
\end{quote}
The HTC and HPC definitions describe distinct operations for problems
which can be cast in the data-parallel paradigm.  However, it is easy
to imagine HPC-type operations included in the {\tt HTCImpl}
component, and the other way around.  For example, {\tt HTCImpl} may
launch multiple data-parallel execution graphs, each with its own {\tt
  HPCImpl} component.  For EPs with multiple CPUs and GPUs, this {\tt
  HPCImpl} component may include HTC-type operations cognizant of data
locality and high bandwidth inter-connection at the EP to further
manage the partitioning of the quantum of the data at each EP and
optimize the workflow for multiple CPU-GPU combination. It is
therefore useful to have {\tt HTCImpl} and {\tt HPCImpl} as distinct
architectural components with the understanding that they may share
parallelization techniques and tool-chain.

\section{Architectural Views}
\label{Sec:ARCH-VIEWS}
Figure~\ref{Fig:GenSolver} shows a view of the conceptual architecture
for a generalized solver build with the architectural components
identified in the previous sections.  In this section we show
different views of this generalized architecture for a variety of
solvers for calibration and imaging models with specialization of the
various components, some of which are further decomposed into
sub-components for re-usability.

We use existing algorithms as case studies to show that the
architecture is foundational in nature.  These core set of algorithms
form the building blocks of an hierarchical domain algorithmic
architecture.  A comprehensive coverage of {\it all} RA algorithms in
use is however out of scope for this paper.  Not all
direction-dependent techniques in use are therefore covered here
(e.g., \cite{TIM_N_RICK_1992_FACETING, WSnapshot,SD-INT}).  While the
architectural framework developed here is designed to accommodate
these and other similar methods, further work is necessary to cast and
verify those algorithms in this framework.

\subsection{Architectural View for Calibration}
\label{Sec:AV-CALIB}
For clarity here we discuss a direction-independent solver for only
the diagonal elements of the Jones matrix (the $G$-Jones solver) and
frequency (the $B$-Jones solver). A solver for $D$-Jones (off-diagonal
elements of the Jones matrix), or a full-Jones solver can be similarly
developed (see e.g., \cite{Bhatnagar_2017_PSC})

The multiplicative gains per interferometer can be expressed as an
outer product of two $2 \times 2$ Jones matrices, one for each antenna
of the interferometer.  The observed data is the modeled as
\begin{equation}
  \vec{V}^{obs}_{ij} = \left[\J_i\otimes \J^\dag_j\right]
  \vec{V}^{Mod}_{ij} + n_{ij}
\end{equation}
The diagonal elements of the Jones matrix represent the complex gain
for the signal from the two nominally orthogonally polarized
receivers, while the off-diagonal elements represent the leakage of
signals from one receiver into another. In terms of the equivalent
Mueller matrix, the corrected data $\vec{V}^c$ is expressed as:
\begin{equation}
  \label{Eq:VCorr}
  \vec{V}^c = M^{-1}_{ij} \vec{V}^{Obs}
\end{equation}
In general, the overall Mueller matrix is separated into a product of
a series of Mueller matrices, each modeling a different part of the
signal chain.  The precise sequences of these matrices is determined
by the signal path and in general the product is non-commutative.  As
a result, in general, effects of the matrices on the left of the
Mueller matrix in the ME being solved must be first corrected for (see
\myhref{https://casadocs.readthedocs.io/en/stable/notebooks/casa-fundamentals.html}{CASA
  Calibration Fundamentals} document for more detailed description).
Equation~\ref{Eq:VCorr} is this application of calibration terms
before solving for the elements of the Jones matrix of interest.

In the following description, the multiplicative antenna-based complex
gain is denoted by $g^{p}$ for polarization $p$, and the super-script
$pp$ represents interferometer based parallel-hand products. $V^{Mod}$
represents the (fixed) true expected visibilities.  Solving
Eq.~\ref{Eq:dM} with this involves first normalizing $\vec{V}^c$ by
the model data $\vec{V}^{Mod}$ to increase the signal coherence length
and averaging the data for the solution intervals $\Delta T_{sol}$ and
$\Delta \nu_{sol}$ along time and frequency axis respectively as
\begin{eqnarray}
  \label{Eq:VMod}
  \vec{V}^{Mod} &=& \DG\F \vec{I}^{Mod}\\
  \label{Eq:VAvg}
  \vec{X}_{ij} &=& \sum_{t}^{\Delta T_{sol}} \sum_\nu^{\Delta
    \nu_{sol}}\vec{V}_{ij}^{c}(t,\nu)\left[\vec{V}^{Mod}_{ij}(t,\nu)\right]^{-1}
\end{eqnarray}
and the $\rchi^2$ for a single polarization is then expressed in terms
of $X_{ij}$ as
\begin{eqnarray}
  \label{Eq:XChisq}
  \rchi^2 &=& \sum\limits_{ij} \vec{R}_{ij}^{{pp}^\dag}
  \left[W_{ij}\right] \vec{R}_{ij}^{pp} \nonumber
\end{eqnarray}
where $\vec{R}_{ij}^{pp} =\vec{X}^{pp}_{ij} - g^p_i g^{p^\dag}_j$ and
$\left[W_{ij}\right]$ is the weight matrix.
\label{Sec:Cal}
\begin{figure*}[ht]
\begin{center}
  \includegraphics [height=3in,width=\textwidth]{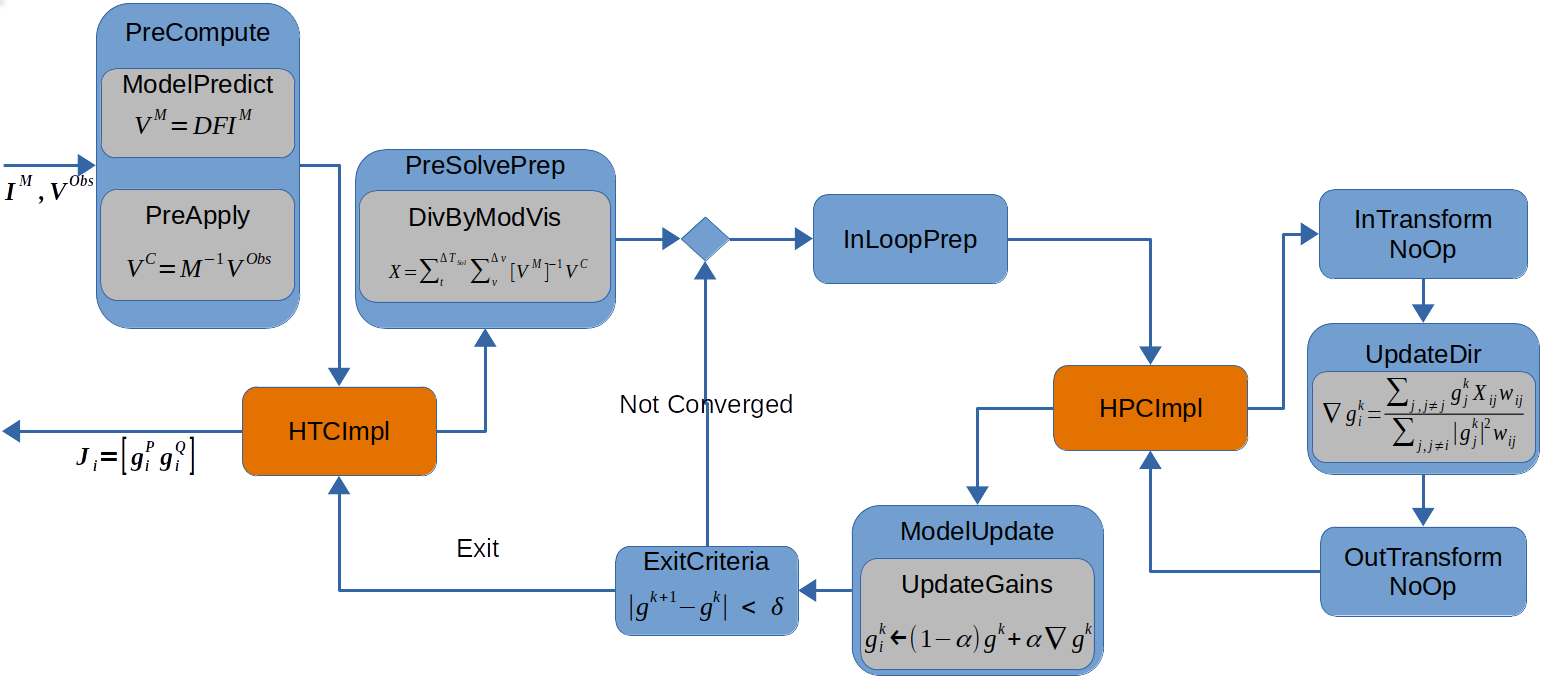}
  \caption {\small A specialization of the Generic Solver of
    Fig.~\ref{Fig:GenSolver} for a simple diagonal G-Jones
    solver. A different specialization of the components shown in gray
    would deliver solvers for other calibration terms.  Inputs are the model image
    ($I^M$) and the uncalibrated data ($V^{Obs}$).  Output is the
    diagonal elements of the Jones matrix (super-scripts $P$ and $Q$
    represent the two orthogonal polarizations).}
  \label{Fig:CalSolver}
\end{center}
\end{figure*}

Equating the derivative $\partial \rchi^2/\partial g^p_i$ to zero
leads to a recursive closed-form solution for $g^p_i$ at iteration $k$,
expressed as:
\begin{eqnarray}
\label{Eq:CalUpdateDir}
\nabla g^{p^{k}}_i &=& \frac{\sum\limits_{j,j\ne i}{} g^{p^k}_jX^{pp}_{ij}w^{pp}_{ij}}{\sum\limits_{j,j\ne
    i}\left|g^{p^k}_j\right|^2 w^{pp}_{ij}}\\
\label{Eq:CalUpdateModel}
  g^{p^{k+1}}_i &=& \left(1-\alpha\right)g_i^{p^k} + \alpha \nabla g_i^{p^{k+1}}
\end{eqnarray}
$\alpha$ here plays the role of the update step size and balances the
contribution of the current solution and the current $\nabla
g^{p^k}_i$. The constraint $\left| \vec{g}^{p^{k+1}} -
\vec{g}^{p^k}\right|< \delta$ is used for the termination criteria
where $\delta$ is the tolerance limit on the incremental change in the
solutions consistent with the estimated antenna sensitivity.
\subsubsection{Component level mapping}
\begin{figure*}[ht]
\begin{center}
  \includegraphics [height=3.4in,width=\textwidth]{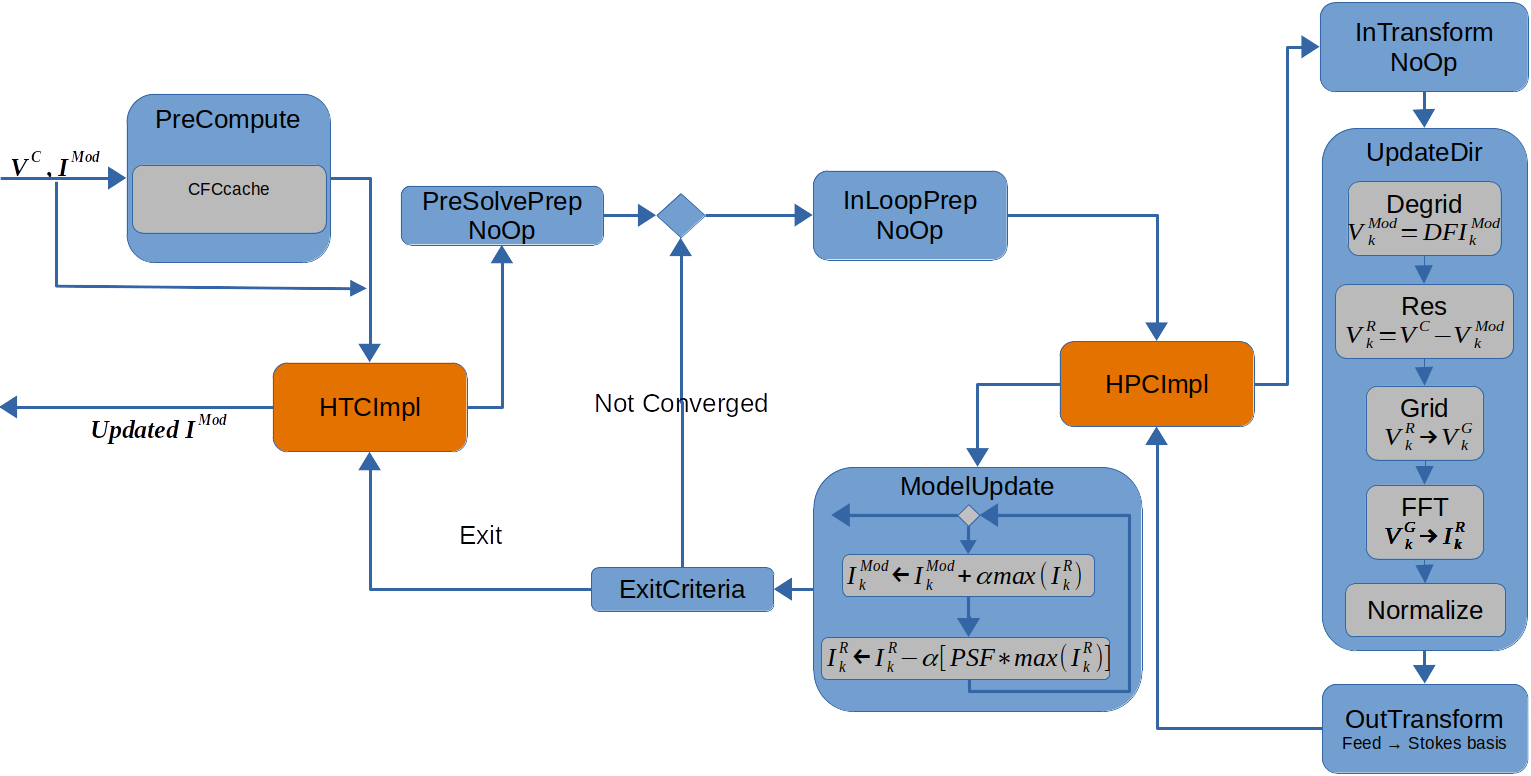}
  \caption {\small A specialization of the Generic Solver of
    Fig.~\ref{Fig:GenSolver} for an iterative
    image solver for single pointing imaging. A different specialization
    of the {\tt ModelUpdate} and {\tt UpdateDir} components would deliver
    other variants of the solver.  The sub-components in gray,
    especially of the {\tt UpdateDir} component are re-usable.}
  \label{Fig:ImSolver}
\end{center}
\end{figure*}

The above steps map on to the generic components described in
Sec.~\ref{Sec:Components} and Fig.~\ref{Fig:GenSolver} as
follows.
\begin{itemize}
\item Equations~\ref{Eq:VCorr} and \ref{Eq:VMod} map on to the {\tt
  PreCompute} component.  For DI solver, these are a one-time
  computation and not dependent on iterations.  This is the model
  prediction and data correction operation which in general requires
  the $\DG$ and the $\F$ operators, which map on to the re-usable
  sub-components {\tt Degrid} and {\tt FFT} discussed in
  Sec.~\ref{Sec:Imaging} and shown in Fig.~\ref{Fig:ImSolver}.
  Simpler implementations for model prediction and correction are
  possible for special cases.

\item Equation~\ref{Eq:VAvg} maps on to the {\tt PreSolvePrep}
  component. The summations over time and frequency can be done
  independently allowing partitioning of the input data.  The solver
  can also be applied on each chunk of data in parallel for the
  solutions in time and frequency on coherence scales of $\Delta
  T_{sol}$ and $\Delta \nu_{sol}$.  This algorithm-level
  parallelization is achieved via dispatch from the {\tt HTCImpl}
  component, or done internally in the component.

\item The {\tt InLoopPrep} component may be a NoOp, or may include
  heuristics or algorithms, for example, for detection and rejection
  of poor solutions and data.

\item Eq.~\ref{Eq:CalUpdateDir} maps on to the {\tt UpdateDir}
  component.  This may also benefit from multi-threaded
  parallelization at the EP (e.g., by using all available CPU cores at
  each EP in the {\tt HPCImpl} component).

\item Eq.~\ref{Eq:CalUpdateModel} maps on to the {\tt ModelUpdate}
  component.

\item Finally, the constraint $\left| \vec{g}^{p^{k+1}} -
  \vec{g}^{p^k} \right| <
  \delta$ maps on to the {\tt ExitCriteria} component.
\end{itemize}
Note that implementations that require numerical evaluation of the
derivative -- e.g. where closed-form equation is not possible or
practical -- are also possible.  Finally, for terms where
transformation and averaging of the data as in Eq.~\ref{Eq:VAvg} is
not possible, the {\it current} solution may be used in the {\tt
  InLoopPrep} component to compute the residual vector
($\vec{R}_{ij}$) along with related modifications in the {\tt
  PreSolvePrep}, {\tt UpdateDir} and {\tt ModelUpdate} components.
\subsection{Architectural View for Imaging}
\label{Sec:Imaging}

Equation~\ref{Eq:Chisq} for imaging can be expressed as:
\begin{equation}
\label{Eq:ImChisq}
\chi^2 = \left[V^C -\DG \F I^{Mod}_k\right]^\dag \W
\left[V^C -\DG \F I^{Mod}_k\right]
\end{equation}
where $\W$ is the weights matrix, and subscript $k$ represents the
iteration number.  With some simple calculus and algebraic
manipulations, it can be shown that $\frac{\partial \rchi^2}{\partial
  I^{Mod}_k} \propto I_k^R$ and
\label{Sec:Im}
\begin{eqnarray}
  \label{Eq:ResIm}
&I_k^R&= \sum_t \sum_\nu \sum_{ij} \F^\dag\G_{ij}\W_{ij}\left[V^C(t,\nu) - V_k^{Mod}(t,\nu)_k\right]_{ij}
\end{eqnarray}
The interpretation of the summations over $t$ and $\nu$ may change
depending on the type of imaging modality (continuum, spectral cube,
multi-term imaging, or snapshot/temporal movie,etc.).  For example, the
summation over $\nu$ for continuum imaging would be an accumulation on
to a single image, while for spectral cube imaging it maps individual
image(s) on to plane(s) of an image cube.  As in the case of the
calibration solver, the summations can be done independently, and each
can be further partitioned depending on the computer resources.

The model update step of numerical optimization goes by the term
``Minor Cycle'' in the RA literature.  It is itself iterative in
nature, and encapsulated here as
\begin{equation}
  \label{Eq:ImUpdate}
  I_{k+1}^{Mod} = \text{ModelUpdate}\left(\frac{\partial \rchi^2}{\partial I^{Mod}}, I_k^{Mod}\right)
\end{equation}
The algorithms used for {\tt ModelUpdate} range from simple ones that
model the sky brightness distribution as a collection of delta
functions (e.g. \cite{Hogbom_Clean}), to complex ones that also
account for variations along time, frequency and polarization axis in
a scale-sensitive manner (e.g. MS-Clean, MSMFS, Asp-Clean).  Details
of these algorithms can be found elsewhere in the literature (e.g.,
see
\myhref{https://casadocs.readthedocs.io/en/stable/notebooks/synthesis\_imaging.html}{CASA
    Imaging Fundamentals}, \cite{MS-Clean}, \cite{MSMFS},
\cite{AspClean}). Mathematically, a modification of Eq.~\ref{Eq:ResIm}
with a straightforward application of the derivative chain rule to
compute the update direction for hidden parameters and using them in
the {\tt ModelUpdate} component covers these algorithms in the same
architectural framework.

The iteration termination criteria is typically a complex heuristic
involving the derivative ($I_k^R$), current model ($I_k^{Mod}$),
number of iterations done ($N_{iter}$), and mechanisms for forced
termination.  We encapsulate this in a function as
\begin{equation}
  \label{Eq:ImExit}
  \text{ExitCriteria} = T(I_k^R, I_k^{Mod}, N_{iter})
\end{equation}
The derivative computation in general is the most expensive step in
most numerical optimization problems.  For imaging the computational
cost of the $\G$ and $\DG$ operators scale with the total number of
independent data points $N_{vis}$ as $\mathcal{O}(N_{vis}\times
C_{Sup})$, where $C_{Sup}$ is the support size of the interpolation
kernels in number of pixels on the visibility grid. In an end-to-end
processing this step may dominate the overall computing cost.  The
{\tt HPCImpl} component for imaging is therefore typically more
complex and much work has been done in the community in general for
making the calculations for it and the {\tt ModelUpdate}
computationally efficient \citep[e.g.,][]{IDG,ngVLA-HPG,Clustered-Asp,
  WAsp}.

\subsubsection{Component level mapping}

The operations described in Sec.~\ref{Sec:Imaging} map on to the blocks
in Sec.~\ref{Sec:Components}, Fig.~\ref{Fig:GenSolver} as
follows.
\begin{itemize}
\item The {\tt PreCompute} component in
  Fig.~\ref{Fig:ImSolver} as a place holder for one-time operations
  (e.g. caching the calculations for the convolutional kernels used in
  the $\DG$ and $\G$ operators).

\item Similarly, the {\tt PreSolvePrep} and {\tt InLoopPrep} component
  is a place holder for operations on the quantum of data that flows
  to the {\tt UpdateDir} component (e.g. data pre-conditioning,
  re-ordering for runtime performance benefits, etc.).

\item Eq.~\ref{Eq:ResIm} maps on to the {\tt UpdateDir}
  component. This involves several computationally distinct operations
  shown in Fig.~\ref{Fig:ImSolver} as reusable sub-components.  The
  combined operation of this component is referred to as the ``Major
  Cycle'' in RA literature (see
  {\tt CASA Imaging Fundamentals} document for more details).

\item The summations in Eq.~\ref{Eq:ResIm} maps on to the {\tt
  OutTransform} component. The summations may be implemented as an
  accumulation (e.g. accumulating the frequency-binned images on a
  single image for continuum imaging) or a gather operation
  (e.g. gathering a range of frequency bins on separate planes of an
  image cube), or a combination thereof.

\item Eq.~\ref{Eq:ImUpdate} maps on to the {\tt ModelUpdate} component
  (referred to as the ``Minor Cycle''in the RA literature).

\item Eq.~\ref{Eq:ImExit} maps on to the {\tt ExitCriteria} component.
  This may contain sophisticated iteration control heuristics.

\item As in calibration, the {\tt HTCImpl} is responsible for the
  coarse-gain (algorithm- or cluster-level) parallelization.  The
  implementation of this will depend on the architecture of computing
  platform and the parallelization strategy used.  {\tt HPCImpl}
  implements the EP-level parallelization (multi-threading, use of
  multiple CPUs and GPUs).
\end{itemize}
\begin{figure*}[ht]
\begin{center}
  \includegraphics [height=4in,width=6in]{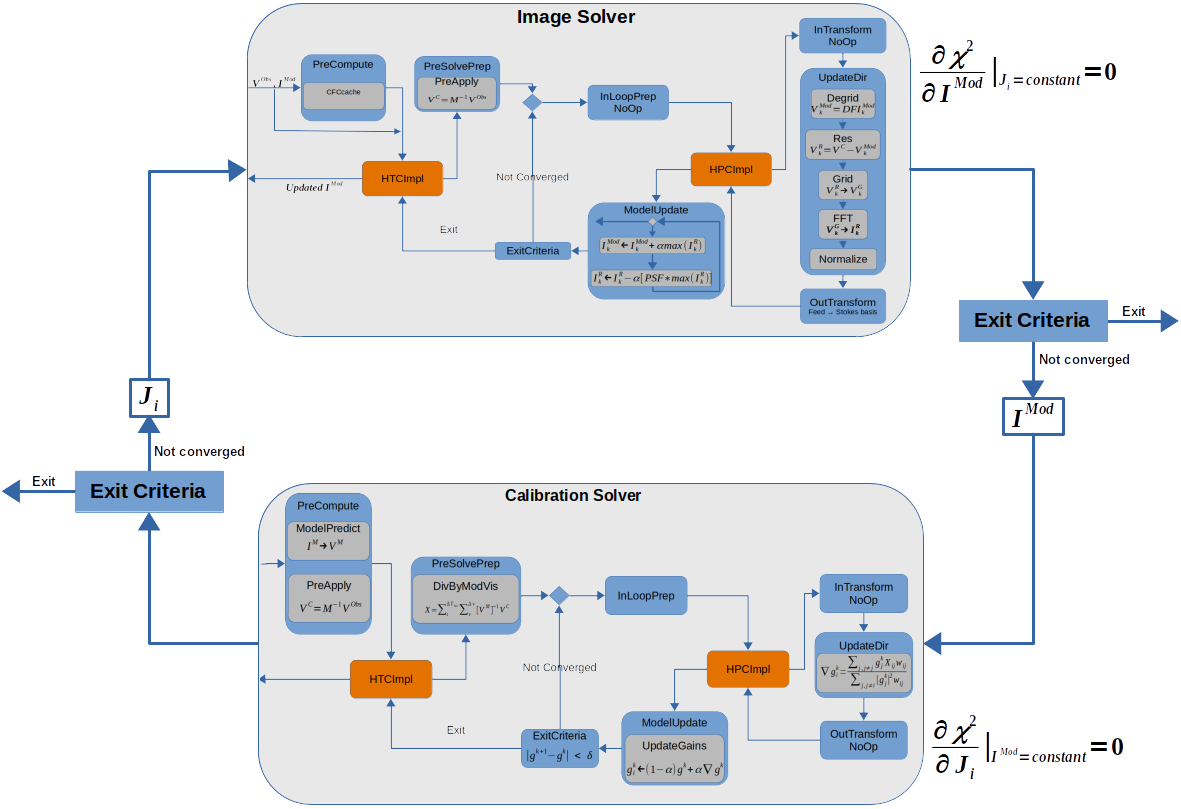}
  \caption {\small Component level view of the SelfCal concept
    shown in Fig.~\ref{Fig:CalIm}.}
  \label{Fig:SelfCal}
\end{center}
\end{figure*}

\section{An example of the Physical Architectural View}
\label{Sec:PhysicalArch}
In this section we give an overview of a physical architectural view
for imaging, covering a variety of wide-band single pointing and
mosaic imaging modalities including some DD corrections (e.g., for the
effects of antenna PB, pointing offsets and non co-planar baselines).
For this we use the C++
\myhref{https://safe.nrao.edu/wiki/pub/Software/Algorithms/WebHome/AWPDesign.pdf}{AWP
  imaging framework} from the
\myhref{https://github.com/ARDG-NRAO/LibRA}{\tt LibRA} project, both
of which are described in more detail in the companion paper on
physical architecture \citep{LibRA}.  This framework was used to build
a highly configurable implementation of the {\tt UpdateDir} component.

\begin{figure}[ht]
\begin{center}
  \includegraphics [height=2in,width=3in]{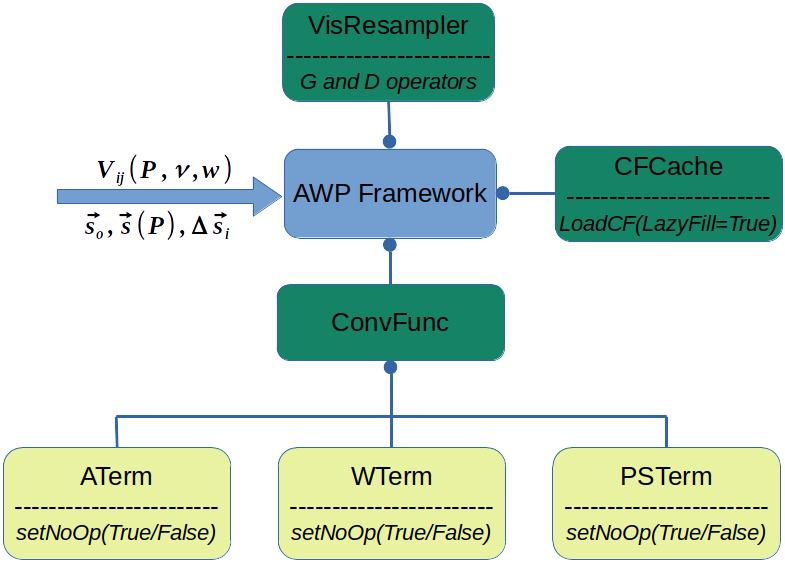}
  \caption {\small A simplified diagram showing the design of the AWP
    framework used in Sec.~\ref{Sec:PhysicalArch} for an example of
    the physical architectural view.  The framework is parameterized
    with the {\tt VisResampler} component which implements the
    re-usable $\G$ and $\DG$ operators, the {\tt CFCache} component
    which implements the in-memory and on-disk models for the cache of
    convolutional kernels $c_{ij}$, and the {\tt ConvFunc} component implements the
    calculations to compute the $c_{ij}$.  The $c_{ij}s$ are further
    parameterized with the {\tt A}, {\tt W} and the {\tt PSTerm}.}
  \label{Fig:AWP-FRAMEWORK}
\end{center}
\end{figure}

\begin{table}[ht]
  \begin{center}
    \begin{tabular}{|c|c|c|c|c|}
      \hline
      \rowcolor{backcolour}
      \hline
      Operation    & {\tt ATerm} & {\tt PSTerm}   & {\tt WTerm} & CF \\
      \hline
      \hline
      \rowcolor{pearl}
      AW-Projection&   ON   &   ON  &   ON   & {\tt PS}$\star${\tt
        A}$\star${\tt W}\\
      \rowcolor{pearl}
                   &        &   OFF &   ON   & {\tt A}$\star${\tt W} \\
      \hline
      A-Projection &   ON   &   ON  &   OFF  & {\tt PS}$\star${\tt A}\\
                   &        &   OFF &   OFF  & {\tt A} \\
      \hline
      \rowcolor{pearl}
      W-Projection &  OFF   &    ON      &  ON    & {\tt
        PS}$\star${\tt W} \\
      \hline
      Standard     &  OFF   &    ON      &  OFF   & {\tt PS} \\
      \hline
    \end{tabular}
    \caption{\small Table of configurations of the AWP imaging
      framework for a variety of projection algorithms. The symbols
      {\tt A}, {\tt PS} and {\tt W} represent the A-term, the
      anti-aliasing Prolate Spheroidal and the W-term functions
      respectively, and $\star$ represents the convolution operator.}
    \label{Tab:AWP-CFSETUP}
  \end{center}
\end{table}
A simplified diagram of the AWP framework is shown in
Fig.~\ref{Fig:AWP-FRAMEWORK}.  The framework is parameterized with the
{\tt VisResampler} object which provides services for the $\G$ ({\tt
  Gridder}) and the $\DG$ ({\tt Degridder}) sub-component, a 2D sparse
Mueller matrix of CFs ($\F\MS{ij}{DD}{}$) used in the $\G$ and $\DG$
operations, and the {\tt CFCache} plugin component to build and manage
a cache of CFs.  The sparse matrix of CFs can also be configured to
include only terms that are significant compared to the telescope
sensitivity.  Configuring these plugin components enables a wide range
of imaging modalities including single pointing, pointed and
on-the-fly mosaic imaging, optionally for full-polarization imaging
with corrections for a variety of DD wide-band, wide-field effects.

The {\tt CFCache} plugin component manages the in-memory and
persistent models of the cache of CFs and can itself be configured as
shown in Tab.~\ref{Tab:AWP-CFSETUP}.  This allows construction of a
variety of DD projection algorithms, including corrections of antenna
pointing offsets.  The calculations for the CFs in the cache can be
triggered in the {\tt PreCompute} component, with the {\tt CFCache}
plugin configured to pre-compute the cache.  It can also be configured
to compute the CFs on-demand for cases where it may be efficient to
compute CFs recurrently on demand.  A combination of these two types
of configurations is also possible which allows growing an existing CF
cache with usage.

For the general case of wide-band full-polarization mosaic imaging
the computations in the {\tt UpdateDir} component can be expressed as:
\begin{equation}
\label{Eq:Appendix-IMEQ}
  I^R(P) =
  \sum\limits_{P,\nu,w,ij}\F^{\dag}\G(c_{ij})W_{ij}\left[V_{ij}\left(P,\nu,w\right)
    - \DG(c_{ij})I^M(P,\nu)\right]
\end{equation}
where $P$ is the array pointing direction towards the part of the sky
being imaged.  The convolution kernels $c_{ij}$ for the interferometer
$ij$ is further parameterized as
\begin{equation}
\label{Eq:CIJ}
c_{ij}\left(P,\nu,w\right) = CF_{ij}(\nu,w,t)~f\left( \Delta \vec{s}_i
- \Delta \vec{s}_j \right)~e^{2\pi\iota\left[\vec{s_\circ} -
    \left(\vec{s}(P) + \Delta \vec{s}_i + \Delta \vec{s}_j
    \right)\right]}
\end{equation}
where $CF_{ij}$ is constructed from ideal kernels and its construction
can be configured as shown in Table~\ref{Tab:AWP-CFSETUP}. While
$CF_{ij}$ can be independent for each data, in practice {\it a} kernel
applies to a set of data points. The rest of the terms in
Eq.~\ref{Eq:CIJ} represents the data-dependent perturbations applied
on-the-fly.  $\vec{s}_{\circ}$ and $\vec{s}(P)$ are the vectors
towards the center of the {\it joint image} and the correlator phase
center for pointing $P$ respectively.  $\Delta \vec{s}_i$ is the
antenna pointing offset for antenna $i$ with respect to the phase
center direction $\vec{s}(P)$. $f$ represents any loss in sensitivity
due to antenna pointing offsets.  It is close to unity for small
offsets with $f(0)=1.0$ and $0 < f <= 1.0$.

$\vec{s_{\circ}}$, $\vec{s}(P)$, and the set of offset vectors
$\left\{\Delta \vec{s}_i\right\}$ are part of input meta-data for the
AWP framework, which enables a number of imaging modalities as
tabulated in Table~\ref{Tab:AWPSETUP-MOSAICKING}.

\begin{table}[ht]
  \begin{center}
    \begin{tabular}{|c|c|c|c|}
      \hline
      \rowcolor{backcolour}
      \hline
      Imaging    &Phase center   & Image center       & Pointing offset  \\
      \rowcolor{backcolour}
      mode       & $\vec{s}(P)$   & $\vec{s}_{\circ}$ & (PO) $\Delta \vec{s}_i$ \\
      \hline
      \hline
      {\bf Single pointing}  &                  &                  &      \\
      No PO correction       &  $\vec{s}_{\circ}$ & $\vec{s}_{\circ}$ &      0\\                     
      With PO correction     &  $\vec{s}_{\circ}$ & $\vec{s}_{\circ}$ &      *\\                     
      \hline
      \rowcolor{pearl}
      {\bf Mosaic imaging}     &                    &                 &     \\
      \rowcolor{pearl}
      No PO correction   &  *                 & $\vec{s}_{\circ}$ &  0 \\
      \rowcolor{pearl}
      With PO correction &  *                 & $\vec{s}_{\circ}$ &     * \\
      \hline
    \end{tabular}
    \caption{\small Table of configurations  for the AWP framework for
      single or multi-pointing imaging, with or without DD corrections
      for antenna pointing offsets.{\tt *} represents value used as-is
      from the input meta data.}
    \label{Tab:AWPSETUP-MOSAICKING}
  \end{center}
\end{table}

Multi-threaded code to drive the {\tt UpdateDir} component and
pre-fetch the data and CFs to overlap computing and i/o operations is
deployed in the {\tt HPCImpl} component.  The implementation of the
{\tt VisResampler} plugin of the AWP framework in the {\tt UpdateDir}
component is implemented in the {\tt Kokkos} framework
\citep{Kokkos-original, Kokkos-latest} for performance-portable
implementation of the $\G$ and $\DG$ operators \citep{ngVLA-HPG}.
This combination of multi-threaded code on the host CPU and
massively-parallel code on the GPU has proven to be scalable, flexible
and portable across variety of EP architectures, each with multiple
CPUs and GPUs in a single large-scale parallel execution.

A variety of frameworks for data distribution and parallel processing
were deployed via the {\tt HTCImpl} component, ranging from simple
Python scripts, generic tools like
\myhref{https://doi.org/10.5281/zenodo.1146013}{\tt gnuparallel}, to
functionally more complex and capable schedulers like {\tt Slurm} and
{\tt HTCondor}.  This allowed use of a variety of computing platforms
ranging from laptop computers, multi-CPU desktop-class computers and
tightly connected clusters, to wide-area distributed clusters and
super-computer class machines.

As a test, we used this software system for imaging a few data sets
which otherwise would take inordinately long to process.  We deployed
the {\tt UpdateDir} component on the Pathways to Throughput (PATh)
computing and the National Research Platform (NRP) computing
facilities of the Open Science Grid in the US, utilizing over 100 GPUs
of four different architectures in parallel.  The {\tt PreCompute}
component was deployed on an in-house cluster using the {\tt Slurm}
scheduler, and the {\tt ModelUpdate} component was deployed on a
single computer with larger memory and faster multi-core CPUs.  One of
the archived data sets we imaged is from deep integration of the
Hubble Ultra Deep Field with the Very Large Array (VLA) in the
  A-array configuration at S-Band using $\sim$2~GHz of the available
  bandwidth and $\sim\!\!100$ hr of on-source integration.  These
  parameters require application of the wide-band AW-Projection
  algorithm \citep{WBAWProjection} to correct for the wide-field
  wide-band instrumental effects and non-coplanarity of the array.
  The sky brightness distribution at these low frequencies is also
  stronger and spatially more complex, requiring the use of the
  MS-Clean algorithm \citep{MS-Clean} to model the sky brightness
  distribution and mapping a field-of-view covering up to the first
  sidelobe of the antenna far-field power pattern. Both of these
  requirements further increases the computing load significantly.  As
  a result, although this data was taken nearly a decade ago, it could
  not be processed due to excessive runtime and computational costs.
With our setup, processing at the rate of $\sim\!\!2$~Terabyte/hour,
convergence was achieved in 10 iterations in about 24~hr of wall-clock
run time.  The resulting image is one of the deepest image with the
VLA with an RMS noise of $\sim\!\!1\mu$~Jy/beam (see
\myhref{https://science.nrao.edu/enews/17.3/index.shtml\#deepimaging}{NRAO
    Newsletter article}, paper in preparation).

\subsection{Hierarchical algorithm development}
The Self-Calibration procedure for a joint solver, discussed in
Sec.~\ref{Sec:GEN-STRUCTURE}, is an example of hierarchically building
higher level algorithms using lower level components. As shown in
Fig.~\ref{Fig:SelfCal} such a joint solver can be built with the image
solver described above and a similar calibration solver. The {\tt
  HTCImpl} component manages the data partitioning and dispatch to EP
nodes.  For a DI-only calibration solver, operations of the {\tt
  PreCompute} component may be combined in the {\tt PreSolvePrep}
component and executed in parallel.  The {\tt HPCImpl} component
manages the execution at the EPs of the {\tt UpdateDir} and {\tt
  ModelUpdate} components.  The results are then assembled for the the
{\tt ExitCriteria} component to determine convergence.  Once
convergence is achieved, the {\tt PreApply} sub-component of the
calibration solver may be used in the {\tt PreSolvePrep} component of
the image solver to use data calibrated for the current calibration
model. The {\tt ExitCriteria} after the image solver or the
calibration may terminate the iterations of this joint solver.

Parallelization is achieved by data partitioning determined by the
solution intervals in time and frequency for the calibration solver.
Any data reordering for runtime optimization of the solvers is done
on-the-fly.  With relatively smaller size of the individual data
partitions compared to the full data, the runtime overhead of data
reordering at the EPs was insignificant.

\section{Discussion and Conclusions}
In this paper we described the algorithms for processing the data from
aperture synthesis telescopes in standard terminology of numerical
optimization and signal processing.  We then developed an algorithmic
architecture based on fundamental mathematical concepts, and show that
the core algorithms for both calibration and imaging are numerical
optimization algorithms which can be expressed in the same
architectural framework.  In this approach, calibration and imaging
algorithms are described as specialized views of the underlying
foundational conceptual architecture.

We make two important points. First, that this architecture is built
on fundamental concepts and is foundational in nature.  Successful
algorithms for traditional computer systems (non-quantum computers) --
currently or in the future -- will require {\it at least} the
architectural components we identify.  Algorithms and their
implementation based on these components will therefore remain
computationally and algorithmically scalable, and adaptable in an
evolving overall computing landscape.  Secondly, the computing
resources required (e.g. memory or FLOP/s footprint) for each of these
components may vary significantly. Implementation and optimization of
each component can therefore be quite different and may require quite
a different tool-chain.  It is easy to imagine that the optimal
technologies used to harvest the computing power at various scales may
also vary from component to component.  The proposed architecture
allows modifications, necessitated by changes in the computing
environment or in the algorithm used for a given component, to remain
confined to that specific component, enabling a modular, mix-and-match
approach between components and the hardware on which they are
deployed.

Replacing the implementation of a component by a different one is also
straightforward, keeping the architecture unchanged.  This not only
makes the implementation easier to evolve and maintain, but also makes
it easier for experts from other domains to contribute to specific
parts of the architecture (e.g., use of special hardware for a
specific component only).  Highly optimized implementation of
individual components may even be available from other communities,
enabling effective external collaborations.  The proposed
architecture is inherently scalable, accommodating the continued
evolution of hardware, software technologies, and domain-specific
algorithms.  It also facilitates collaboration with experts from other
domains, a capability that is increasingly important in the emerging
era of interdisciplinary scientific computing.

As an example, we also present a physical architectural view for an
image solver based on the conceptual architecture.  For this we use
the AWP framework written in C++ for the {\tt UpdateDir} architectural
component.  This framework is parameterized with plug-in components
for the $\G$ and $\DG$ operators, the convolutional kernels used in
these operators, and the {\tt CFCache} plugin to manage the in-memory
and on-disk models of a cache of these kernels.  All of these can be
configured in a variety of ways to realize a wide range of algorithms
for different imaging modalities.  The $\G$ and the $\DG$ operators
are themselves implemented in the {\tt Kokkos} framework
\citep{Kokkos-latest} for performance portable implementation for use
on a variety of massively parallel hardware like GPUs and CPUs with
large number of cores (e.g. the ARM architecture). With
multi-threading in the {\tt HPCImpl} component for high-performance
computing on the host CPU, this design has proved to be flexible and
performance-portable and could be easily deployed on a heterogeneous
network of remote nodes with a variety of architectures. The {\tt
  ModelUpdate} component was implemented to enable a range of
image-plane modeling algorithms from a library of algorithms from the
\myhref{https://github.com/ARDG-NRAO/LibRA}{\tt LibRA} project
(publicly available since 2022), which is a re-engineered fork of the
CASA code base \citep{CASA}.  The {\tt HTCImpl} component for
high-throughput computing was implemented with a variety of
parallelization frameworks on multiple parallel computing platforms
ranging from multi-CPU desktop-class computers to super computer class
distributed and tightly connected clusters.

The configurability of the AWP framework as shown in
Tables~\ref{Tab:AWP-CFSETUP} and \ref{Tab:AWPSETUP-MOSAICKING} allows
a wide range of algorithms for the {\tt UpdateDir} component, while
the configurability of the {\tt CFCache} plugin allows optimization of
the run-time performance and memory management.  The {\tt ModelUpdate}
component is independently configured to invoke the algorithms of
choice, and it could be optimized independently.  In an accompanying
paper \citep{LibRA}, we discuss in more detail the physical
architecture with some results and runtime scaling on a few different
parallel computing platforms.

The capability to implement the various components quite independently
of each other leads to an overall implementation that is efficient and
flexible in utilizing a range of available computing platforms.  This
also enabled effective collaborations with experts from other domains
for the various components which together constitute a complex
computing stack.  For example, the {\tt VisResampler} plug-in for the
AWP framework was developed by in-house experts in collaboration with
the {\tt Kokkos} group and industry R\&D groups
\citep{ngVLA-HPG,Kokkos-latest}.  The {\tt HTCondor}-based
implementation of the {\tt HTCImpl} was developed by in-house experts
in collaboration with the staff at the
\myhref{https://chtc.cs.wisc.edu/}{Center for High Throughput
  Computing (CHTC)} and the \myhref{https://www.sdsc.edu}{San Diego
  Supercomputer Center}. This effectively verifies key architectural
goals of efficiency, flexibility and scalability in terms of using
computing resource, human expertise, and in enabling a wide range of
scientific functionality in an easily configurable implementation.

Here we also note that all this was possible, from conceptualization
to deployment and application to real-life problems, due to the
effective participation from different, very diverse set of experts
from multiple modestly-sized groups, each with a concentration of
expertise in specific parts of the computing stack.  Each group could
focus on a small part of the architecture without the impractical need
of full-system expertise first.  This seems to both, reduce the cost
and increase the speed of development in an increasingly complex and
rapidly evolving overall computing landscape (hardware, software
tool-chain, and domain algorithms \citep{PlentyOfRoomAtTheTop}).
These considerations are of critical importance for the viability of
the next-generation of radio telescopes.  These features together also
support our conclusion that the mathematical formulation and the
resulting architecture described here is foundational in nature.  As a
future goal, expressing other existing algorithms and new algorithms
in this framework without the need for architectural change will
independently verify its foundational nature.

\begin{acknowledgements}
  We thank the members of the Algorithms R\&D Group of the National
  Radio Astronomy Observatory (NRAO) for many useful impromptu
  discussions the over the years.  We also thank Neeraj Gupta (IUCAA,
  Pune), George Moellebrock (NRAO, Socorro), Kumar Golap (NRAO,
  Socorro) and Mark Whitehead (NRAO, Charlottesville) for helpful
  discussions and the anonymous referee for useful suggestions and
  corrections.  The Very Large Array (VLA) telescope is operated by
  the NRAO which is a facility of the National Science Foundation
  operated under cooperative agreement by Associated Universities,
  Inc.
\end{acknowledgements}
 \software{{\tt CASACore} \citep{casacore}, {{\tt CASA} \citep{CASA}},
   {\tt LibRA} \citep[][{\small \tt
       https://github.com/ARDG-NRAO/LibRA}]{LibRA}, {\tt Kokkos}
   \citep{Kokkos-latest}, {\tt HPG} Library \citep{ngVLA-HPG}, {\tt
     WCSLIB} \citep{wcslib_ascl, wcslib_paper}}.

\bibliographystyle{aasjournalv7}
\bibliography{AlgoArch}

\end{document}